\documentclass{aastex63}

\usepackage{color,soul}
\usepackage{lineno}
\received{January, 2022}
\revised{XXX, 2022}
\accepted{7 June 2022}%\today}
\submitjournal{ApJ}

\shorttitle{The isotropic $\gamma$-ray emission above 100 GeV}
\shortauthors{de Menezes et al.}

\graphicspath{{./}{figures/}}
\begin{document}

\title{The isotropic $\gamma$-ray emission above 100 GeV: where do very high energy $\gamma$ rays come from?}

\correspondingauthor{Raniere de Menezes}
\email{Contact e-mail: raniere.m.menezes@gmail.com}

\author[0000-0001-5489-4925]{Raniere de Menezes}
\affiliation{Lehrstuhl f\"ur Astronomie, Universit\"at W\"urzburg, Emil-Fischer-Strasse 31, 97074 W\"urzburg, Germany}
\affiliation{Universidade de S\~ao Paulo, Departamento de Astronomia, Rua do Mat\~ao, 1226, S\~ao Paulo, SP 05508-090, Brazil}

\author{Raffaele D'Abrusco}
\affiliation{Center for Astrophysics | Harvard \& Smithsonian, 60 Garden Street, Cambridge, MA 20138, USA}

\author{Francesco Massaro}
\affiliation{Dipartimento di Fisica, Universit\`a degli Studi di Torino, via Pietro Giuria 1, I-10125 Torino, Italy}
\affiliation{Istituto Nazionale di Fisica Nucleare, Sezione di Torino, I-10125 Torino, Italy}
\affiliation{INAF-Osservatorio Astrofisico di Torino, via Osservatorio 20, 10025 Pino Torinese, Italy}
\affiliation{Consorzio Interuniversitario per la Fisica Spaziale (CIFS), via Pietro Giuria 1, I-10125, Torino, Italy}

\author{Sara Buson}
\affiliation{Lehrstuhl f\"ur Astronomie, Universit\"at W\"urzburg, Emil-Fischer-Strasse 31, 97074 W\"urzburg, Germany}

\begin{abstract}

Astrophysical sources of very high energy (VHE; $>100$ GeV) $\gamma$ rays are rare, since GeV and TeV photons can be only emitted in extreme circumstances involving interactions of relativistic particles with local radiation and magnetic fields. In the context of the \textit{Fermi} Large Area Telescope (LAT), only a few sources are known to be VHE emitters, where the largest fraction belongs to the rarest class of active galactic nuclei: the blazars. In this work, we explore \textit{Fermi}-LAT data for energies $>100$ GeV and Galactic latitudes $b > |50^{\circ}|$ in order to probe the origin of the extragalactic isotropic $\gamma$-ray emission. Since the production of such VHE photons requires very specific astrophysical conditions, we would expect that the majority of the VHE photons from the isotropic $\gamma$-ray emission originate from blazars or other extreme objects like star-forming galaxies, $\gamma$-ray bursts, and radio galaxies, and that the detection of a single VHE photon at the adopted Galactic latitudes would be enough to unambiguously trace the presence of such a counterpart. Our results suggest that blazars are, by far, the dominant class of source above 100 GeV, although they account for only $22.8^{+4.5}_{-4.1}$\% of the extragalactic VHE photons. The remaining $77^{+4.1}_{-4.5}\%$ of the VHE photons still have an unknown origin.

\end{abstract}

\keywords{methods: statistical --- galaxies: active --- gamma rays: general}

\section{Introduction}
\label{sec:intro}

Emitting photons with energies greater than 100 GeV is challenging. In nature, such photons are typically produced under extreme conditions surrounding compact astrophysical sources \citep{Rieger2013_TeV_Astronomy}, such as pulsars or black holes, where charged particles can be accelerated up to relativistic energies and hence emit very high energy (VHE, $>100$ GeV) $\gamma$ rays via inverse Compton (IC) scattering and/or curvature radiation \citep{sturrock1971model,caraveo2014gamma,blandford2019relativistic}. Above $\sim 50$ GeV, only a few hundred $\gamma$-ray sources are significantly detected \citep{wakely2008tevcat,ackermann2016_2FHL} with the \textit{Fermi} Large Area Telescope (LAT) and ground-based observatories, like the High Energy Stereoscopic System \citep{funk2004_HESS}, the Very Energetic Radiation Imaging Telescope Array System \citep{weekes2002_veritas}, and the Major Atmospheric Gamma Imaging Cherenkov Telescopes \citep{baixeras2003magic}.

As listed in TeVCat\footnote{\url{http://tevcat2.uchicago.edu/}} \citep{wakely2008tevcat} and the Second \textit{Fermi}-LAT Catalog of High-Energy Sources\footnote{\url{https://fermi.gsfc.nasa.gov/ssc/data/access/lat/2FHL/}} \citep[2FHL;][]{ackermann2016_2FHL}, the extragalactic VHE sky is dominated by blazars, which are radio-loud active galactic nuclei with a relativistic jet directed toward the observer \citep{urry1995unified,blandford2019relativistic}. Blazars are known for their multifrequency variable emission and are divided mainly into BL Lacs and Flat Spectrum Radio Quasars, which differ mostly by the absence or presence of broad emission lines ($> 5$ \AA) from an accretion disk in their optical spectra and a flat radio spectrum \citep{massaro2009_BZCat1}. On the other hand, most of the VHE Galactic sources are associated to supernova remnants and pulsar wind nebulae \citep{ackermann2016_2FHL}, i.e., sources where effective acceleration of particles in magnetic fields is followed by their interactions with the surrounding gas and radiation fields \citep{reynolds2012magnetic}, thus emitting $\gamma$ rays. Other important contributions to the VHE sky come from Milky Way's interstellar medium, where $\gamma$ rays are produced via cosmic ray interactions with gas and photon fields of star forming regions \citep{ackermann2012_fermi_diffuse_iem}; and from the Fermi bubbles, which are two large structures extending up to nearly $\pm 50^{\circ}$ in Galactic latitude, for which the origin is still under debate \citep{su2010_Fermi_bubbles_original,ackermann2014_Fermi_bubbles,miller2016_Fermi_bubbles}.

The large positional uncertainty of $\gamma$-ray observations, especially below 1 GeV, makes the association of $\gamma$-ray sources with their low-energy counterparts a challenging task \citep{massaro2016extragalactic}, as several astrophysical objects can lie within their positional reconstruction confidence regions \citep{massaro2016extragalactic,abdollahi2020_4FGL,abdollahi2022incremental}. Therefore, the astrophysical community relies on statistical association methods, where observed quantities are used to compute an association probability between the $\gamma$-ray source and its counterpart candidates \citep{sutherland1992likelihood,ackermann2011_2LAC,abdollahi2020_4FGL,deMenezes2020_physical_assoc}.

At high Galactic latitudes (i.e., $b \gtrsim |50^{\circ}|$), where $\gamma$-ray emission from interstellar medium and Fermi bubbles is almost negligible, most of the photons with more than 100 GeV could in principle be associated with blazars, especially with the BL Lac type \citep{massaro2013BllacVHE}, which typically present harder $\gamma$-ray spectra. At such energies, the 68\% containment radius for a single photon detection with \textit{Fermi}-LAT is $\sim 0.1^{\circ}$ \citep{atwood2009LAT}, and therefore this single detection could unveil the presence of a blazar. Blazars indeed account for nearly 80\% of the sources listed in the third data release of the \textit{Fermi}-LAT fourth source catalog \citep[4FGL-DR3;][]{abdollahi2020_4FGL,abdollahi2022incremental} for such Galactic latitudes and are prone to emit VHE photons via inverse Compton processes (especially BL Lac objects).

In this work, we investigate the origin of the isotropic $\gamma$-ray emission for energies $> 100$ GeV and Galactic latitudes $b >|50^{\circ}|$. We want to test whether this isotropic VHE emission can be interpreted as coming from a population (or many populations) of sources. In other words, we want to test if the detection of a single VHE photon can be attributed to a potential astrophysical counterpart. At such energies and Galactic latitudes, blazars are by far the dominant population of sources in the sky and we would expect them to have a major role in the extragalactic isotropic $\gamma$-ray emission \citep{wakely2008tevcat,ackermann2016_2FHL}. Furthermore, recent theoretical results \citep{roth2021diffuse} suggest that star-forming galaxies may account for most (if not all) isotropic $\gamma$-ray emission in the \textit{Fermi}-LAT energy band. As an additional piece of information, we know that VHE photons cannot travel through cosmological distances due to $\gamma$-$\gamma$ interactions with the extragalactic background light \citep[EBL;][]{finke2010_EBL,Dominguez2011_EBL,Gilmore2012_EBL,dominguez2013_gamma-ray_horizon}, implying that the VHE sky is mainly dominated by photons coming from redshifts $z \lesssim 1$. At such depth, the counterparts of the majority of the $> 100$ GeV photons must be relatively nearby sources accessible to lower-energy surveys.

Results achieved in this work will have direct applications to the science developed with the upcoming Cherenkov Telescope Array \citep[aka CTA;][]{actis2011_CTA,acharya2013introducing}. This paper is organized as follows. In \S \ref{sec:Analysis}, we describe the selections we apply to \textit{Fermi}-LAT data and in \S \ref{sec:hunting} we describe the analysis methods. In \S \ref{sec:Results} we detail the results for the adopted sky regions and, finally, we summarize and conclude our findings in \S \ref{sec:discussion}.

\section{Data selection}
\label{sec:Analysis}

This work uses data collected by the \textit{Fermi}-LAT during its first 12.5 years of operation, i.e., from August 4th 2008 to February 10th 2021, with energies above 100 GeV. We select only \textit{Fermi}-LAT events belonging to the {\small \texttt{SOURCEVETO}} class \citep[\texttt{evclass} $= 2048$;][]{bruel2018fermi}, which is one of the cleanest event classes in Pass 8 \citep{atwood2013pass} and is recommended for studies of diffuse emission that require low levels of cosmic-ray contamination\footnote{Additional information about all types of event classes can be found in the Fermi Science Support Center webpage: \url{https://fermi.gsfc.nasa.gov/ssc/data/analysis/documentation/Cicerone/Cicerone_Data/LAT_DP.html}}. We consider events converted in the front or back sections of the telescope tracker (\texttt{evtype} $= 3$), and in two regions of interest (ROIs) above $|50^{\circ}|$ in Galactic latitude, hereinafter referred as north and south ROIs. At such latitudes, we avoid 90\% of the predicted Galactic interstellar $\gamma$-ray emission at energies around 100 GeV, and discard nearly 95\% of the Galactic sources listed in 4FGL-DR3. To reduce the contamination from the Earth limb, a maximum zenith angle cut of $105^{\circ}$ is applied, and we apply the standard good time interval filter {\small \texttt{DATA\_QUAL}} $>0$ and the recommended instrument science configuration {\small \texttt{LAT\_CONFIG}} $== 1$.

For the main analysis, no exposure maps, livetime cubes or source maps are computed. We aim to analyze each event of our dataset individually. To estimate the Galactic contribution at such high latitudes and energies, we compare the predicted number of photons from four distinct Galactic models, namely \texttt{gll\_iem\_v07}, \texttt{gll\_iem\_v06}, \texttt{gll\_iem\_v05} and \texttt{gal\_2yearp7v6\_v0} in the energy range 100--500 GeV for both ROIs. This energy range has been chosen because the Galactic models prior to \texttt{gll\_iem\_v07} do not reach TeV energies. For each model, we compute the total number of VHE photons by directly integrating them over the time, energy, and solid angle intervals adopted here. These Galactic models are compositions of several multiwavelength templates mapping the gas and dust in the Milky Way and differ in several aspects, as in the angular resolution of the gas distributions, and the gas and dust density/velocity profiles over the Galaxy\footnote{The several components used in the Galactic model \texttt{gll\_iem\_v07} can be found on a document in this website: \url{https://fermi.gsfc.nasa.gov/ssc/data/access/lat/BackgroundModels.html}. Unfortunately, this work was not published in a peer-reviewed journal.}. We select the upper and lower values for the total number of photons estimated with these Galactic models as the systematic error band (in the 100--500 GeV energy range) and consider the predicted number of photons from \texttt{gll\_iem\_v07} to be the most accurate value among all the models, since it is the latest model released to date. When we extrapolate these values for energies $> 500$ GeV (based on the energy distribution of photons in our dataset) and consider the Poissonian statistical error on the total number of photons, we get a total of $315 \pm 39$ (sys) $\pm 18$ (stat) and $239 \pm 30$ (sys) $\pm 15$ (stat) photons predicted by the Galactic model for the northern and southern ROIs, respectively.

Other major background contributors are the Fermi Bubbles. However, as shown in the next section, only the Southern bubble is a significant source of VHE photons on the analysis performed here and we remove it from our ROI. Our two ROIs are centered in the North and South Galactic poles and include 1165 and 794 VHE photons, respectively. This substantial difference in the number of photons is caused mainly by the uneven coverage of the sky performed by \textit{Fermi}-LAT, which has a deeper exposure towards the Northern Galactic hemisphere \citep{ackermann2012_LAT_exposure_map} and by the presence of Mkn 421 in the northern ROI.

\begin{figure}
    \centering
    \includegraphics[width=\linewidth]{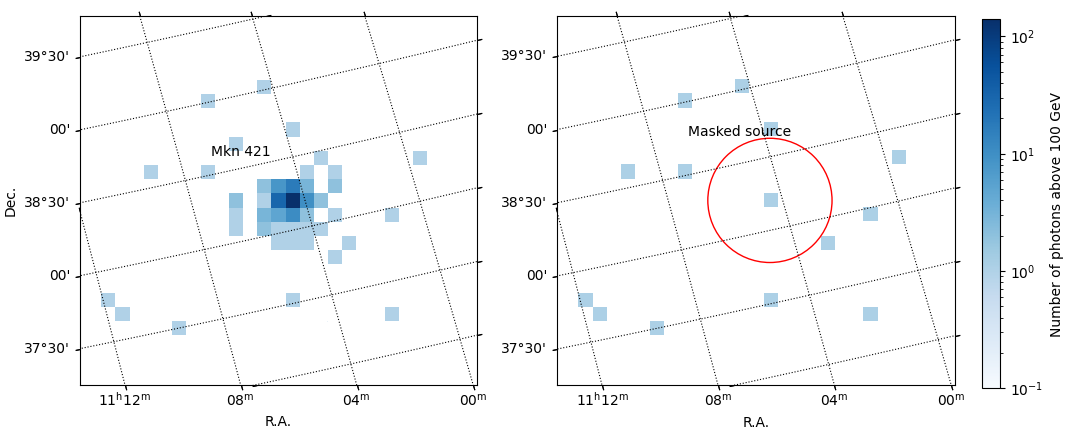}
    \caption{Mask applied to Mkn 421. Here, all VHE photons within a circle of radius $0.5^{\circ}$ and centered on the position of Mkn 421 are mapped to a single photon in the center of the circle. Blue squares have a size of $0.1^{\circ} \times 0.1^{\circ}$ and the color scale represents photons with $> 100$ GeV.}
    \label{fig:Mkn421_masked}
\end{figure}

\section{Methods}
\label{sec:hunting}

Several $\gamma$-ray blazars are relatively bright VHE sources. The VHE photons that they emit appear as clusters of events in the LAT data. In order to avoid to erroneously associate these clustered photons with multiple sources, we start our analysis by crossmatching the VHE photons with 4FGL-DR3 and masking those photon clusters coincident with 4FGL-DR3 sources, such that, to the aims of our analysis, they are represented by a single photon centered on the position of the $\gamma$-ray counterpart listed in 4FGL-DR3. From now on, we refer to our sample of photons after masking as ``equivalent photons", since some of them may represent clusters of photons. In Figure \ref{fig:Mkn421_masked} we show the effect of applying the mask to Mkn 421, where all VHE photons lying within a radius of $0.5^{\circ}$ from the $\gamma$-ray source center are substituted by a single equivalent photon. Adopting this strategy, we aim to be very conservative in excluding any clustering of VHE photons related to 4FGL-DR3: as can be seen in Figure \ref{fig:Mkn421_masked}, for strong VHE sources, like Mkn 421, a small percentage ($\sim 3\%$) of VHE photons can be spread over a radius of $0.3^{\circ}\sim 0.4^{\circ}$. The total number of equivalent photons in the northern ROI drops to 743 after masking 58 clusters, while in the Southern ROI it drops to 656 after masking 41 clusters.

\begin{figure}
    \centering
    \includegraphics[width=\linewidth]{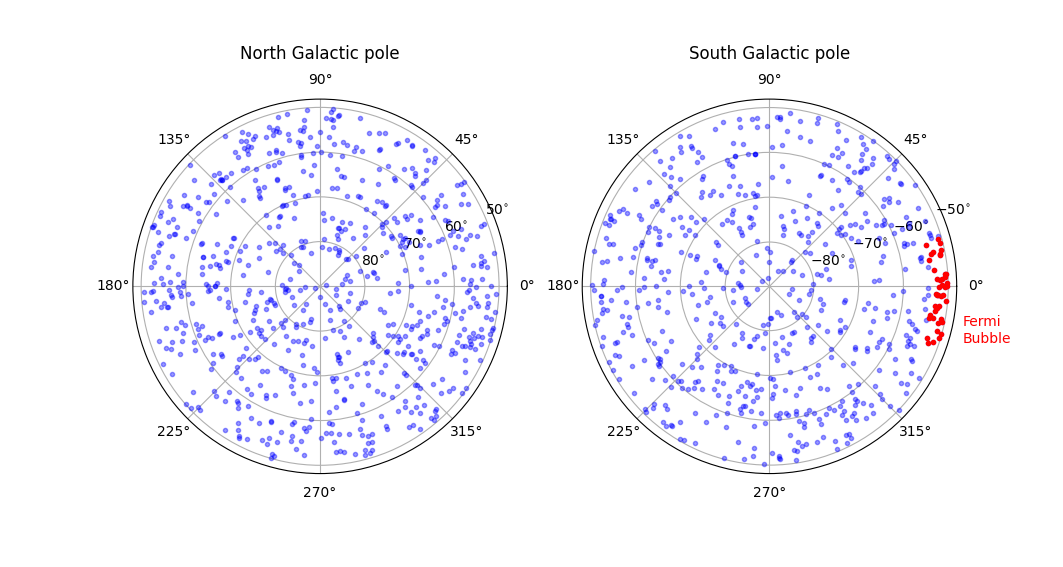}
    \caption{Equivalent photons above 100 GeV in the northern and southern ROIs after masking for photon clusters coincident with 4FGL sources. Red dots represent equivalent photons coincident with the Southern Fermi Bubble and are excluded from the rest of the analysis. In principle, we assume that each blue dot can be associated to a blazar.}
    \label{fig:Fermi_bubble}
\end{figure}

\begin{figure}
    \centering
    \includegraphics[scale=0.4]{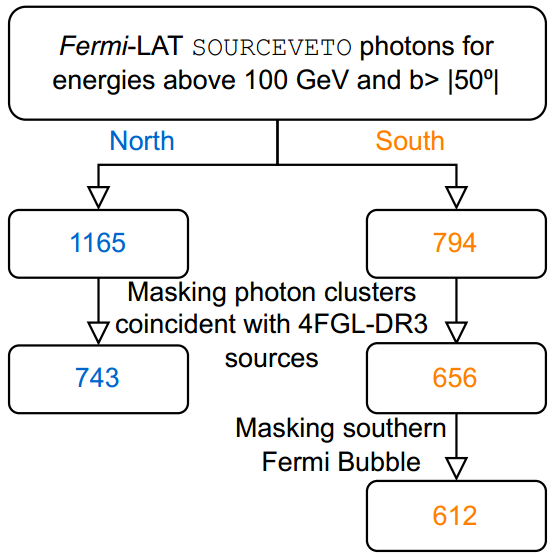}
    \caption{Flowchart with the cuts applied to \textit{Fermi}-LAT data. Blue numbers represent the VHE photons in the northern ROI, while orange numbers represent those in the southern ROI.}
    \label{fig:flow}
\end{figure}

The spatial distribution of equivalent photons after masking photon clusters coincident with 4FGL-DR3 sources is shown in Figure \ref{fig:Fermi_bubble}, where the panels are centered at the North and South Galactic poles. Differently from the Northern Fermi Bubble, which barely reaches $b = 50^{\circ}$ in Galactic latitude and does not contaminate our ROI, the Southern Fermi Bubble extends down to $b = -54^{\circ}$ and is a significant source of background for our analysis. We therefore discard all equivalent photons spatially coincident with the bubble, as shown by the red dots in Figure \ref{fig:Fermi_bubble}. The chosen edges are $b > -54^{\circ}$, in Galactic latitudes, and $-22^{\circ} < l < 16^{\circ}$ in Galactic longitude, in accordance with the bubble edges available in \cite{sarkar2019possible}. The number of discarded equivalent photons is 44, reducing the total number of equivalent photons in the Southern ROI to 612. In Figure \ref{fig:flow} we summarize all cuts applied to \textit{Fermi}-LAT data for both ROIs.

After masking, we assume that each VHE equivalent photon is a potential $\gamma$-ray source and crossmatch them with several catalogs of astrophysical counterparts. Our main catalog is comprised of 5648 blazars collected from Roma-BZCat \citep{massaro20155_BZCAT}, 4FGL-DR3, and identified in optical spectroscopic campaigns \citep{massaro2013unveiling,paggi2014optical,crespo2016opticalCampVI,marchesini2019optical,pena2019optical}. Most of the blazars listed in this catalog have a confirmed optical spectrum (nearly 100\% in the case of BL Lacs) or are confirmed $\gamma$-ray emitters. The impact of adopting other catalogs of $\gamma$-ray sources is discussed in \S \ref{sec:results_other_cats}. To test if these blazars indeed contribute to the VHE isotropic emission, we repeat the matching process for 5000 lists of mock VHE equivalent photons, generated by displacing the original positions of the VHE equivalent photons by a random value between $0^{\circ}$ and $5^{\circ}$ in a random direction of the sky (see \S \ref{sec:Results}).

\begin{figure}
    \centering
    \includegraphics{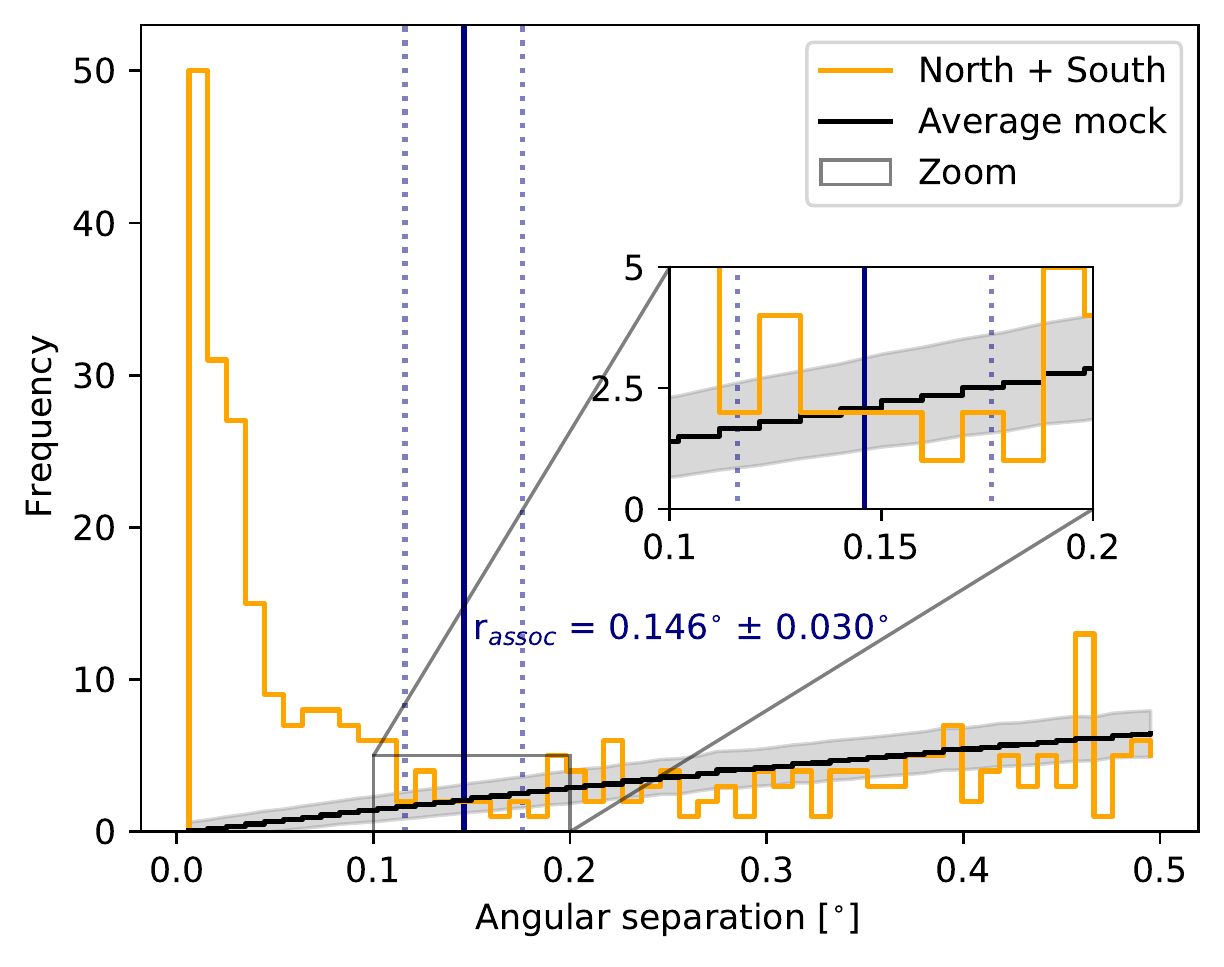}
    \caption{Distribution of angular separation between VHE equivalent photons and the blazars in our main catalog (orange line). The black line represents the average of the same distribution computed for 5000 mock equivalent photon lists and the gray shadow is the standard deviation of this distribution. We define the association radius as $r_{assoc} = 0.146^{\circ}$, which is the median value for the intersection between the orange line and each one of the 5000 angular separation distributions for the mock lists of equivalent photons. The error in $r_{assoc}$ is taken as the standard deviation of these intersection points and has a value of $0.030^{\circ}$. Throughout this work, we use the rounded value of the association radius $r_{assoc} = 0.15^{\circ} \pm 0.03^{\circ}$. Inset: zoom in the region around $r_{assoc}$.}
    \label{fig:separation}
\end{figure}

In order to define the optimal association radius, we first perform the crossmatches between the VHE equivalent photons and the sources listed in our main catalog considering a quite large constant association radius $r = 0.5^{\circ}$, and then compare the distribution of angular separation between the real and mock equivalent photon lists. From Figure \ref{fig:separation}, we define the optimal association radius to be $r_{assoc} \approx 0.15^{\circ}\pm0.03^{\circ}$, beyond which the associations are consistent with noise. This procedure has been successfully adopted in several works searching for the counterparts of $\gamma$-ray sources in radio, infrared and optical bands \citep[more details can be found in][]{massaro2013ApJS..207....4M,massaro2014AJ....148...66M,massaro2014ApJS..213....3M,giroletti2016A&A...588A.141G}. We assume $r_{assoc}$ to be constant over the adopted energy range, which is reasonable, given that the \textit{Fermi}-LAT containment radius for photons above 100 GeV is nearly constant \citep[see section 2 in][]{abdo2011_fermi_Sun}. If instead we use other catalogs to define the optimal association radius (as those discussed in \S \ref{sec:results_other_cats}), we find very similar values for $r_{assoc}$, all around $0.15^{\circ}$ and consistent within the $\sigma_r$ error region.

%In other works \citep{tramacere2013gamma,campana2015application,campana2016application} where authors adopt the Density Based Spatial Clustering of Applications with Noise (aka DBSCAN) and the Minimum Spanning Tree (aka MST) methods to identify $\gamma$-ray sources, they typically use an association radius of $0.3^{\circ}$ for $\gamma$-rays $\gtrsim 10$ GeV. When we test this association radius with our VHE photons (i.e., $> 100$ GeV), the total number of associations increase as much as the noise does, thus we conclude that choosing an association radius of $0.3^{\circ}$ does not improve our analysis.

\section{Results}
\label{sec:Results}

\begin{figure}
    \centering
    \includegraphics{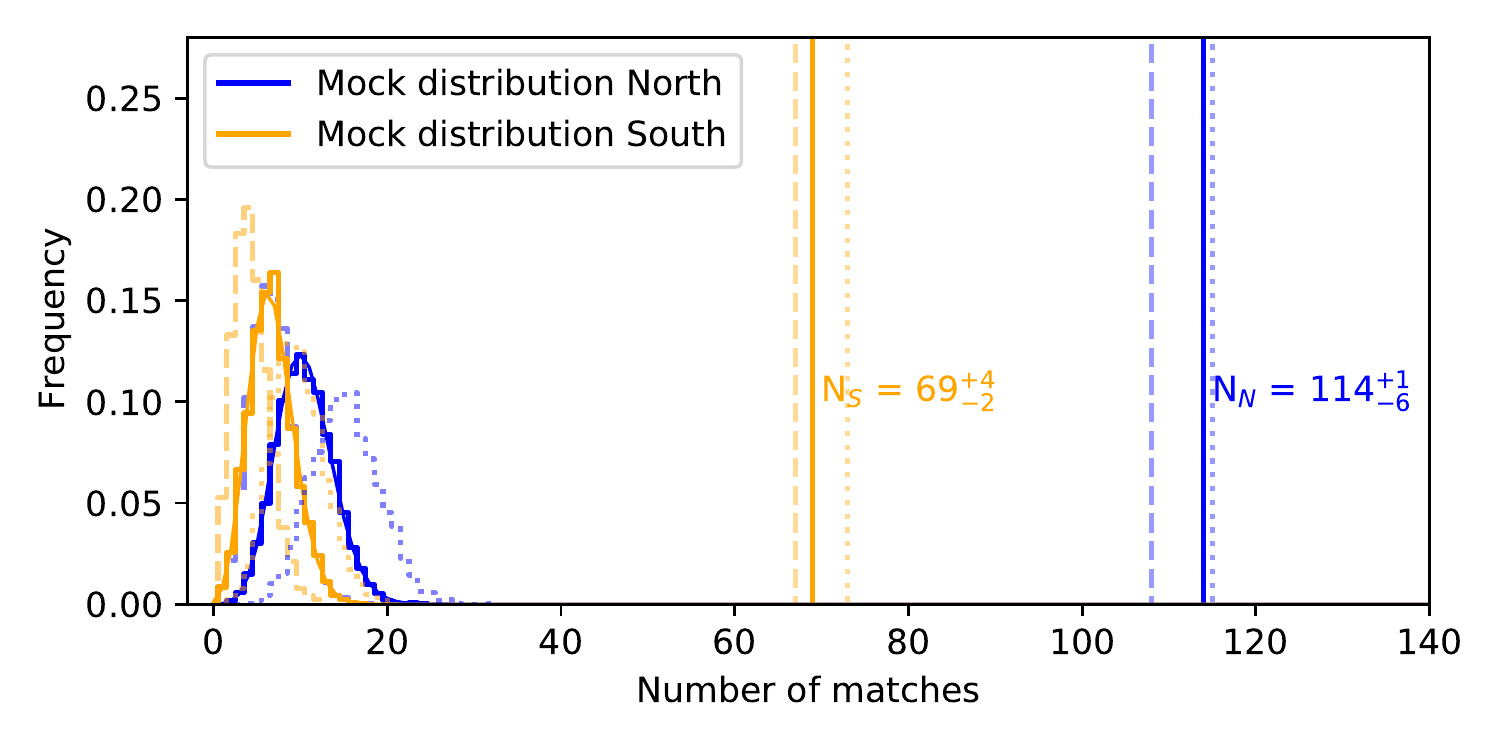}
    \caption{The total number of equivalent photons associated with blazars are $114^{+1}_{-6}$ and $69^{+4}_{-2}$ for the northern and southern ROIs, respectively, considering $r_{assoc} = 0.15^{\circ}\pm0.03^{\circ}$, as shown in Table \ref{tab:fractions}. The orange (South) and blue (North) distributions represent the typical number of matches for 5000 mock lists of VHE equivalent photons (solid lines are computed considering $r_{assoc} = 0.15^{\circ}$, while the dotted and dashed lines are computed for $r_{assoc} = 0.15^{\circ}+\sigma_r$ and $r_{assoc} = 0.15^{\circ}-\sigma_r$, respectively). We fit each distribution with a Poissonian function, which ensures that blazars significantly contribute to the VHE sky at the $32.4\sigma$ level in the North and $24.3\sigma$ in the South (combined significance of $40.3\sigma$).}
    \label{fig:matches}
\end{figure}

The results for both ROIs are shown in Figure \ref{fig:matches}. For the northern ROI (hereinafter represented in blue), we find that $114^{+1}_{-6}$ equivalent photons (after applying the masks previously discussed and considering the $1\, \sigma$ error in $r_{assoc}$) are associated with $114^{+1}_{-6}$ blazars. The expected number of random associations derived from our mock lists of VHE equivalent photons is $10^{+5}_{-3}$. For the southern ROI (hereinafter represented in orange), we have $69^{+4}_{-2}$ equivalent photons associated with $69^{+4}_{-2}$ blazars, with an expected number of random matches of $7^{+2}_{-3}$. In this figure, we fit the distributions of mock matches with a Poissonian function and the results indicate that blazars indeed account for a significant fraction of the VHE equivalent photons detected by \textit{Fermi}-LAT, with an association level of confidence of $32.4\sigma$ in the North and $24.3\sigma$ in the South. However, most of the VHE equivalent photons still lack an astrophysical counterpart listed in the catalogs adopted in this work.

The basic statistics of our findings are shown in Table \ref{tab:fractions}. We observe that $15.3^{+1.5}_{-2.0}$\% of the VHE equivalent photons in the northern ROI are associated with blazars, while for the southern ROI this fraction is $11.3^{+1.8}_{-1.5}$\%. Furthermore, only $10.5^{+1.0}_{-1.1}\%$ of all blazars scattered over both ROIs, as listed in our main catalog, have a corresponding VHE $\gamma$-ray counterpart detected with \textit{Fermi}-LAT, most of which ($\sim 70\%$) are BL Lac objects, although this class of source only makes 28\% of our main catalog. These results tell us that $86.5^{+1.6}_{-1.4}\%$ of \textit{Fermi}-LAT VHE equivalent photons at high Galactic latitudes are unlikely to be originated in blazars. Moreover, as the fraction of associated blazars is smaller than the fraction of associated equivalent photons, it is unlikely that the high number of non-associations is due to incompleteness in our blazar catalog (more details in \S \ref{sec:discussion}). 

As discussed in \S \ref{sec:Analysis}, the expected number of Galactic VHE photons for both ROIs is $315 \pm 39_{sys} \pm 18_{stat}$ in the North and $239 \pm 30_{sys} \pm 15_{stat}$ in the South, which correspond to $42 \pm 8\%_{total}$ and $39 \pm 7\%_{total}$ of all VHE equivalent photons available in both ROIs, respectively. This means that after taking the Galaxy and the blazars into account, roughly 50\% of the observed VHE equivalent photons still have unknown origin (assuming that none of the equivalent photons coincident with blazars are spurious or have Galactic origin). This is equivalent to saying that only $22.8 ^{+4.5}_{-4.1}$\% of the extragalactic photons have a blazar counterpart and that the remaining $77^{+4.1}_{-4.5}\%$ of the extragalactic photons have unknown origin. A possible origin for a fraction of these photons, however, can be due to spurious signals induced by cosmic rays at the detector level (see \S \ref{sec:discussion}).

\begin{table}
    \centering
    \begin{tabular}{l|c|c|c}
        & North & South & North + South\\
        \hline
        Number of equiv. photons & 743 & 612 & 1355\\
        Number of blazars & 977 & 764 & 1741\\
        Matches equiv. photons/blazars  & $114^{+1}_{-6}$ & $69^{+4}_{-2}$ & $183^{+5}_{-8}$\\
        Median matches mock lists &  $10^{+5}_{-3}$ & $7^{+2}_{-3}$  & $17^{+7}_{-6}$\\
        Expected number of Galactic photons & $315 \pm 39_{sys} \pm 18_{stat}$ & $239\pm 30_{sys} \pm 15_{stat}$ & $554 \pm 69_{sys} \pm 23_{stat}$ \\
        Fraction of associated equiv. photons & $15.3^{+1.5}_{-2.0}\%$ & $11.3^{+1.8}_{-1.5}\%$ & $13.5^{+1.4}_{-1.6}\%$\\
        Fraction of associated blazars & $11.7^{+1.1}_{-1.6}\%$ & $9.0^{+1.5}_{-1.2}\%$ & $10.5^{+1.0}_{-1.1}\%$\\
        Fraction of photons from the Galaxy & $42\pm8\%_{total}$ & $39\pm7\%_{total}$ & $41\pm7\%_{total}$ \\
        Frac. of extragal. photons assoc. to blazars & $26.6^{+4.7}_{-5.7}$\% & $18.5^{+4.5}_{-4.0}$\% & $22.8^{+4.5}_{-4.1}$\% 
        
    \end{tabular}
    \caption{Results from crossmatches. Columns represent the northern and southern ROIs individually and summed. The errors are estimated from the uncertainty in $r_{assoc}$ (i.e. computing the number of matches for $r_{assoc}$, $r_{assoc}+\sigma_r$, and $r_{assoc}-\sigma_r$) and Galactic models. For the uncertainties in the fractions, we also take into account the binomial formula (also known as Wald method). We see that only $22.8^{+4.5}_{-4.1}$\% of the extragalactic VHE equivalent photons have a blazar counterpart.}
    \label{tab:fractions}
\end{table}

\subsection{Clustering of VHE equivalent photons}

\begin{figure}
    \centering
    \includegraphics[scale=1]{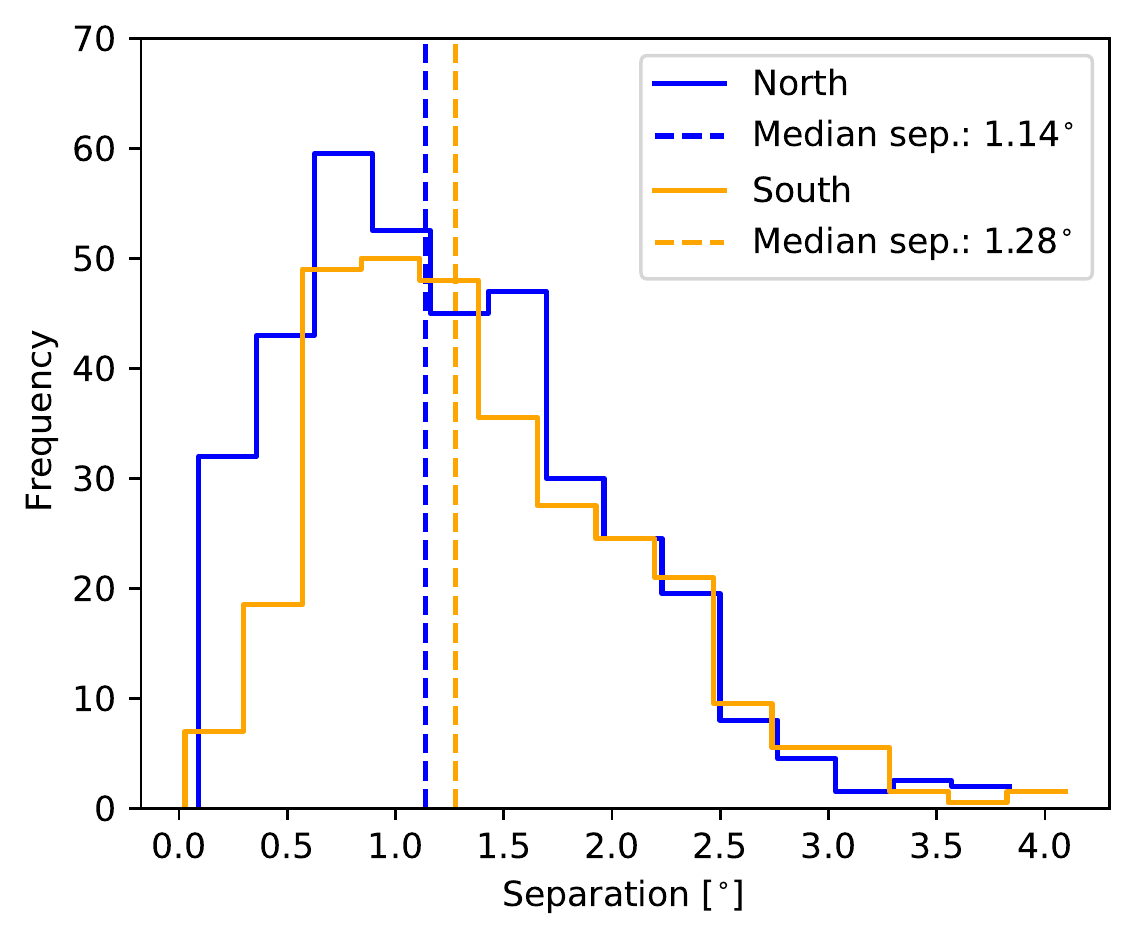}
    \caption{Distribution of the number of equivalent photon pairs in terms of angular separation obtained by self-crossmatching our list of equivalent photons. VHE equivalent photons are typically $1.14^{\circ}$ (North) or $1.28^{\circ}$ (South) away from each other.}
    \label{fig:self_cross}
\end{figure}

There are basically no clusters of VHE equivalent photons within an angular separation of $2\times r_{assoc}$. The exceptions are 8 pairs of equivalent photons in the southern and 15 pairs in the northern ROIs, and none of them match with the blazars in our catalog. These values are compatible with what is expected by chance ($15 \pm 5$ for the north and $10 \pm 4$ for the south) when we crossmatch the lists of VHE equivalent photons from both ROIs with 5000 random distributions of VHE photons scattered over the same areas of the sky and with the same size of our original sample of equivalent VHE photons. In Figure \ref{fig:self_cross}, we show the distribution of angular separation between each pair of equivalent photons in our ROIs and their closest neighbors. We see that most VHE equivalent photons are truly isolated in the sky, being typically more than $1^{\circ}$ away from each other.

\subsection{Repeating the analysis for other catalogs}
\label{sec:results_other_cats}

We repeat the crossmatching process with other catalogs of candidate $\gamma$-ray emitters, again considering the whole sample of VHE equivalent photons and $r_{assoc} = 0.15^{\circ}\pm0.03^{\circ}$ as the association radius (see \S \ref{sec:hunting}). For FRICAT and FRIICAT, two catalogs of Fanaroff-Riley (FR) radio galaxies \citep{capetti2017fricat,capetti2017friicat}, we find only 3 matches with FR I galaxies in the Northern ROI. Similarly, if we adopt a catalog containing only the non-blazars in 4FGL-DR3, we find only 8 matches in the North (three radio galaxies, one active galactic nucleus and three unidentified $\gamma$-ray sources, aka UGSs) and 7 in the South (one radio galaxy, one starburst galaxy and five UGSs). 

\begin{figure}
    \centering
    \includegraphics{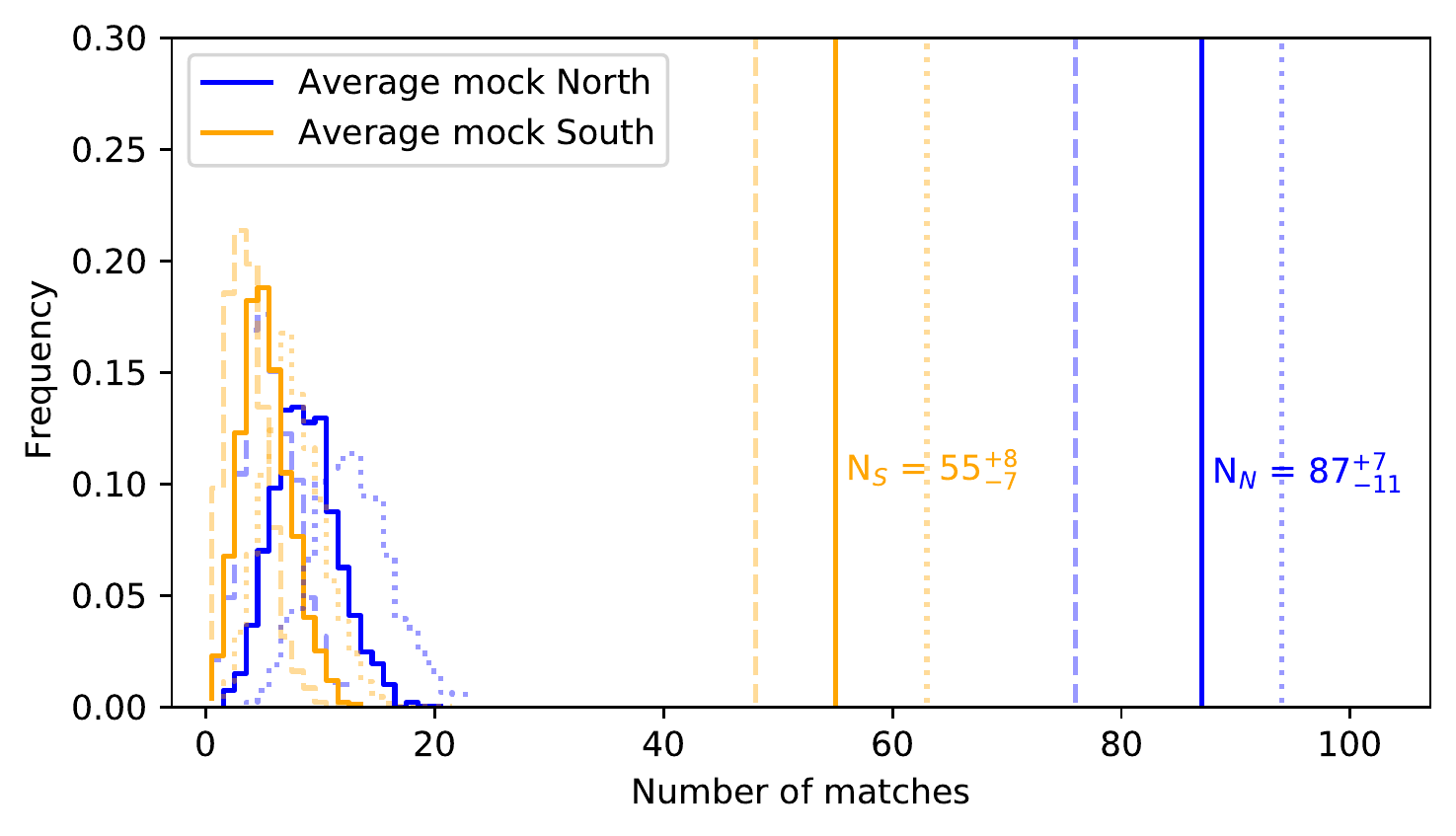}
    \caption{Total number of associations adopting WISECATS as the catalog of blazar counterparts. The blue and orange distributions represent the typical number of matches for 5000 mock lists of VHE equivalent photons considering $r_{assoc}=0.15^{\circ}$ (solid line), $r_{assoc}=0.15^{\circ}+\sigma_r$ (dotted line), and $r_{assoc}=0.15^{\circ}-\sigma_r$ (dashed line). We see that the number of associations is smaller ($87^{+7}_{-11}$ in the North and $55^{+8}_{-7}$ in the South) than those shown in Figure \ref{fig:matches}. We therefore conclude that WISECATS do not perform better than our main catalog in spotting VHE blazar candidates.}
    \label{fig:matches_WIB_KDE}
\end{figure}

When we use the combined sample of WIBRaLS and KDEBLLACS (hereinafter WISECATS), which are two catalogs of $\gamma$-ray blazar candidates \citep{dAbrusco2019WIBRaLS,demenezes2019optical} with 15121 sources in total (2468 sources in the Northern and 1929 in the Southern Galactic ROIs, i.e., far more sources than in our main catalog), the total number of matches is only $142^{+15}_{-18}$ ($87^{+7}_{-11}$ equivalent photons in the North and $55^{+8}_{-7}$ equivalent photons in the South), which is significantly smaller than what we have with our main catalog (see Figure \ref{fig:matches}). Out of these matches, $36^{+2}_{-3}$ associations are not common to our main catalog. By taking these matches into account, the fraction of associated equivalent photons shown in Table \ref{tab:fractions} increases from $15.3^{+1.5}_{-2.0}$\% to $18.0^{+1.5}_{-1.9}$\% in the North and from $11.3^{+1.8}_{-1.5}$\% to $13.9^{+1.7}_{-1.3}$\% in the South (after subtracting the Galactic photons, the fraction of associated equivalent photons becomes $27.3^{+4.8}_{-4.3}\%$ over both ROIs, against the $22.8^{+4.5}_{-4.1}\%$ value reported in Table \ref{tab:fractions}). The results for WISECATS are shown in Figure \ref{fig:matches_WIB_KDE} together with the distribution of matches obtained from 5000 mock lists of equivalent photons. In order to properly compute these noise distributions, we first select WISECATS blazar candidates with $b > |50^{\circ}|$ and, out of those, we randomly select 959 candidates in the North and 730 in the South, so that they match the number of blazars that we have for both ROIs in the main catalog (see Table \ref{tab:fractions}). We then generate 10 lists of mock VHE equivalent photons (as discussed in \S \ref{sec:hunting}) and crossmatch them with the randomly selected blazar candidates. After repeating this process 500 times, we obtain the final noise distributions with a total of 5000 lists of mock VHE equivalent photons. We therefore conclude that WISECATS, although presenting a higher completeness, do not present better results than our main catalog in spotting VHE blazar candidates.

We also test other catalogs of possible $\gamma$-ray emitters, like the Catalogue of Extreme \& High Synchrotron Peaked Blazars \citep[3HSP;][]{Chang2019_3HSP} and the Candidate Gamma-Ray Blazar Survey Source Catalog \citep[CGRaBS;][]{healey2008cgrabs}. For 3HSP we have a total of $129^{+4}_{-6}$ matches for both ROIs, but only $11\pm 1$ of them are not common to our main catalog, while for CGRaBS we find $19\pm 1$ matches in the North and $10\pm 1$ in the South, all of them being common to our main catalog. For the Chandra ACIS Survey for X-Ray AGN in Nearby Galaxies \citep[CHANSNGCAT;][]{she2017chandra}, we find only $6\pm 1$ matches in the North and $3^{+0}_{-1}$ in the South, while for the Australia Telescope National Facility Pulsar Catalogue \citep[ATNF;][]{Manchester2005_ATNF} we find only 1 match for each ROI. Additionally, we find only 3 matches for the transient \textit{Fermi} LAT Second Gamma-Ray Burst Catalog \citep[FERMILGRB;][]{ajello2019decade} and 80 matches for the \textit{Fermi} GBM Burst Catalog \citep[FERMIGBRST;][]{gruber2014fermi,von2014second,bhat2016third,von2020fourth}, this time assuming a larger and fixed association radius of $r_{assoc} = 0.5^{\circ}$ instead of $r_{assoc} = 0.15^{\circ}\pm0.03^{\circ}$. The arrival times of the VHE photons, however, are not consistent with the trigger times of the $\gamma$-ray bursts, meaning that a correlation is unlikely. Among these alternative catalogs, 3HSP and WISECATS lead the significance on the number of sources associated with VHE equivalent photons. We summarize the matches obtained with these alternative catalogs and the combined significance of both ROIs for each one of them in Table \ref{tab:all_cats}.

\begin{table}
    \centering
    \begin{tabular}{l|c|c|c|c}
        Catalog & Matches in the North & Matches in the South & N$^{\circ}$ of sources in both ROIs & Combined sig. ($\sigma$)\\
        \hline
        3HSP & $79^{+1}_{-5}$ & $50^{+3}_{-1}$ & 770 & 35.3\\
        ATNF & $1\pm 0$ & $1\pm 0$ & 92 & 0.5\\
        CGRaBS & $19\pm 1$ & $10\pm 1$ & 506 & 11.5 \\
        CHANSNGCAT & $6\pm 1$ & $3^{+0}_{-1}$ & 473 & 2.8 \\
        FERMIGBRST* & 44 & 36 & 758 & $\sim 0$\\
        FERMILGRB*  & 1 & 2 & 49 & $\sim 0$\\
        FRICAT & $3\pm 0$ & $0\pm 0$ & 120 & 1.3\\
        FRIICAT & $0\pm 0$ & $0\pm 0$ & 72 & 0\\
        WISECATS & $87^{+7}_{-11}$ & $55^{+8}_{-7}$ & 4617 & 28.1\\

    \end{tabular}
    \caption{List of alternative catalogs and their total number of matches with the VHE equivalent photons considering $r_{assoc} = 0.15^{\circ}\pm 0.03^{\circ}$. The last column gives the combined significance of the matches for both ROIs. For the catalogs tagged with ``*", the adopted association radius is fixed at $r_{assoc} = 0.5^{\circ}$.}
    \label{tab:all_cats}
\end{table}

The situation somewhat changes for the Veron Catalog of Quasars \& AGN \citep[VERONCAT;][]{veron2010catalogue}, where $577^{+98}_ {-110}$ VHE equivalent photons ($452^{+80}_{-82}$ in the North and $125^{+18}_{-28}$ in the South) are associated with $1153^{+465}_{-386}$ active nuclei (mostly quasar candidates), and for the Million Quasars Catalogue \citep[MILLIQUAS;][]{Flesch2019_MILLIQUAS}, where 1270 VHE equivalent photons ($734^{+8}_{-22}$ in the North and $536^{+36}_{-80}$ in the South) are associated with $7453^{+3189}_{-2625}$ quasar candidates. These several multiple matches are, of course, a consequence of the very high number of sources in these catalogs, preventing us from precisely tracking the counterparts of the VHE equivalent photons. We also analyze the optical color diagrams of the matches achieved with VERONCAT and MILLIQUAS using the optical colors and magnitudes available in these catalogs, but they do not show any peculiar property over the general populations of VERONCAT and MILLIQUAS sources. We also tested if the VHE equivalent photons match with a list of 35 active galactic nuclei known to host ultrafast outflows \citep[see this list in][]{lat2021gamma_ufos}, but we found only a couple of matches, leading to an association significance below $1.5\sigma$.

In a recent work by \cite{roth2021diffuse}, the authors found that most (or possibly all) of the isotropic $\gamma$-ray emission could be explained by star-forming galaxies. The authors also argue that for energies $\sim 1$ TeV, the major contributors to the isotropic $\gamma$-ray emission are nearby ($z \sim 0.1$) star-forming galaxies. To test this scenario, we collected 64 nearby ($z_{max} \sim 0.06$) star-forming galaxies from \cite{ackermann2012_starforming}, which have unambiguous ongoing star formation, selected based on the presence of dense molecular clouds spotted by the HCN survey \citep{gao2004hcn}. Our results show that only a couple of these star-forming galaxies match with the VHE equivalent photons, giving an association significance of $\sim 1\sigma$, and suggesting that nearby ($z \leq 0.06$) star-forming galaxies (as selected here) may not significantly contribute to the isotropic $\gamma$-ray emission at energies $> 100$ GeV (assuming that the $\gamma$-ray emission of star-forming galaxies is correlated with their infrared emission collected in the HCN survey).

%%%%%%%%%%%%%%%%%%%%%%%%%%%%%%%%%%%%%%%%%%%%%%%%%%%%%%%%%%%%%%%%%%%%%%%%%%

\section{Discussion and conclusions}
\label{sec:discussion}

For energies $>100$ GeV and Galactic latitudes $b > |50^{\circ}|$, we would expect the extragalactic isotropic $\gamma$-ray emission to be dominated by blazars or other potential VHE sources, like star-forming and radio galaxies. It is unreasonable to assume that these VHE equivalent photons originate from Galactic VHE objects, like pulsar wind nebulae and supernova remnants, as out of 62 such objects listed in 4FGL-DR3, only two have  $b \sim 32^{\circ}$ (both in the Large Magellanic Cloud), and the rest has $b < |10^{\circ}|$ \citep{abdollahi2020_4FGL,abdollahi2022incremental}. Above $|50^{\circ}|$ in Galactic latitude, the majority of VHE $\gamma$-ray emitters must be extragalactic. In fact, in the third data release of 4FGL, only 1.4\% of the $\gamma$-ray sources with $b > |50^{\circ}|$ are associated with Galactic counterparts (i.e. 18 pulsars and one binary system), while 79\% of them are associated with blazars.

After subtracting the expected number of Galactic VHE photons from our sample, we are left with $\sim 800$ equivalent photons (North + South). In the most optimistic scenario, where all blazar associations are real, only $22.8^{+4.5}_{-4.1}$\% of these equivalent photons can be associated with blazars, leaving $\gtrsim 75\%$ of the isotropic $\gamma$-ray emission unexplained. Given the incompleteness of our main catalog (described in \S \ref{sec:hunting}), the real fraction of associations can be higher, however, if incompleteness is really the problem, in \S \ref{sec:results_other_cats} we should observe more matches with WISECATS than with our main catalog, as this catalog has a much higher completeness for BL Lac objects, which are the main candidates for VHE emission \citep{massaro2013BllacVHE}. Our results therefore tend to disfavor unknown blazars as the sources of the majority of VHE equivalent photons from the extragalactic isotropic $\gamma$-ray emission. Instead, we could speculate other origins for them, like for instance the scattering of the EBL by ultra-high energy cosmic rays or the scattering of soft photons in the Galactic halo that are not appropriately accounted for in the \texttt{gll\_iem\_v07} model. The electromagnetic cascades resulting from such interactions could in principle create a truly diffuse isotropic $\gamma$-ray emission \citep{ackermann2015spectrum}. Other possibilities for this emission could rise from the interaction of cosmic rays with the Solar radiation field \citep{orlando2008gamma} and Solar System debris \citep{moskalenko2009isotropic}, which are expected to contribute to the isotropic emission at some still unknown level.

%The intensity attributed to the IGRB is observation dependent because more sensitive instruments with deeper exposures can extract fainter extragalactic sources, whereas the total EGB intensity (assuming complete subtraction of all Galactic emissions) is the fundamental quantity.

Furthermore, at the detector level, \textit{Fermi}-LAT can eventually misinterpret a cosmic-ray induced event with a $\gamma$-ray, then introducing a spurious signal to the real data. We reduce the impact of this systematic problem by adopting the \texttt{SOURCEVETO} event class (see \S \ref{sec:Analysis}) in our analysis, which is one of the most stringent \textit{Fermi}-LAT event reconstruction classes in discriminating cosmic rays from $\gamma$ rays, even though it is not 100\% effective. In the P8R3 \texttt{SOURCEVETO} \textit{Fermi}-LAT data adopted in this work, the intensity of cosmic-ray induced events above 100 GeV is roughly constant, with a value of $\sim 10^{-6}$ MeV cm$^{-2}$ s$^{-1}$ sr$^{-1}$ (\textit{Fermi}-LAT Collaboration, 2022, in preparation), which can account for $\sim 1\%$ of the events at 100 GeV, $\sim 3\%$ at 300 GeV, and less than 10\% at 500 GeV. Since $\sim 90\%$ of the photons in our sample (before applying the masks) have energies below 300 GeV, the final level of contamination by cosmic rays is $<< 10\%$. These levels of contamination do not significantly affect the fraction of blazar associations shown in Table \ref{tab:fractions} since, except for a handful of blazars, all matches in our analysis (before masking) have at least one VHE photon below 300 GeV, where the contamination is minimal.

Previous works on the isotropic $\gamma$-ray emission \citep[see e.g.][]{harding2012models,singal2012flux,ajello2015origin} also found that blazars cannot account for all of the extragalactic isotropic $\gamma$-ray emission, although none of them used a method similar to the method adopted here. In \cite{ackermann2015spectrum}, the authors found that roughly half of the total extragalactic $\gamma$-ray emission at 100 GeV come from blazars which, in our work, would be equivalent to measuring the total number of photons in both ROIs (before masking) that are originated by blazars. For a matter of comparison, in our work the fraction of photons (before masking) coincident with blazars is $43.1\%$, in reasonable agreement with \cite{ackermann2015spectrum}, given the significant differences in our analyses. Furthermore, \citep{roth2021diffuse} recently found that a significant fraction of the isotropic $\gamma$-ray emission at $\sim 1$ TeV can be attributed to the emission of nearby ($z \approx 0.1$) star-forming galaxies using a theoretical approach. In this scenario, the VHE emission of star-forming galaxies is generated by cosmic rays accelerated in supernova remnants interacting with the interstellar medium. Our investigation of an observed list of low redshift ($z \leq 0.06$) star-forming galaxies described in \S \ref{sec:results_other_cats}, disfavors the hypothesis that star-forming galaxies are the origin/counterparts of our list of VHE equivalent photons, at least in the very local Universe ($z \leq 0.06$).

In this work we used \textit{Fermi}-LAT observations to investigate whether the $>100$ GeV extragalactic isotropic $\gamma$-ray emission above $b = |50^{\circ}|$ can be explained by blazars. Our results suggest that blazars are not enough to explain this emission and are summarized below.

\begin{enumerate}
    \item Only $15.3^{+1.5}_{-2.0}$\% of all equivalent photons in the northern and $11.3^{+1.8}_{-1.5}$\% in the southern ROIs are associated with blazars. After subtracting the Galactic photons from our sample, the fraction of extragalactic equivalent photons associated with blazars is $22.8^{+4.5}_{-4.1}$\%. If additionally to the main catalog we also consider WISECATS, this fraction slightly increases to $27.3^{+4.8}_{-4.3}\%$. This result suggests that the detection of a single VHE $\gamma$ ray at $b > |50^{\circ}|$ does not unambiguously lead to the detection of a blazar counterpart.
    \item Based on the adopted catalogs, we found that $\gtrsim 75\%$ of the extragalactic isotropic $\gamma$-ray emission above $100$ GeV and $b > |50^{\circ}|$ has no clear origin. 
    \item Almost 70\% of the matches are with BL Lac objects, which are sources known to typically present harder $\gamma$-ray spectra if compared to other blazars like Flat Spectrum Radio Quasars.
    \item The tests we performed with 5000 mock catalogs ensure that the association of blazars with VHE equivalent photons is not by chance and that blazars do have a contribution, although small, in the VHE sky. It is unlikely that the observed small fraction of associations is due to the incompleteness of our catalog. The fact that we have more blazars than photons in both ROIs and also that we performed the same analysis with alternative, and more complete catalogs, rule out this possibility. 
    
\end{enumerate}

With the advent of the CTA \citep{actis2011_CTA,acharya2013introducing} in the upcoming years, new possibilities for investigating VHE sources are expected, helping in the quest of unveiling the origin of the extragalactic isotropic $\gamma$-ray emission.

\acknowledgments

We thank Andrew Strong, Dario Gasparrini, Elena Orlando, Markus Ackermann, Melissa Pesce-Rollins, Michela Negro, Philippe Bruel, and the anonymous referee for the constructive comments allowing us to significantly improve the manuscript. R.M. and S.B. acknowledge financial support by the European Research Council for the ERC Starting grant MessMapp, under contract no. 949555. R.D'A. is supported by NASA contract NAS8-03060 (Chandra X-ray Center). This work is supported by the ``Departments of Excellence 2018 - 2022’’ Grant awarded by the Italian Ministry of Education, University and Research (MIUR) (L. 232/2016). This research has made use of resources provided by the Ministry of Education, Universities and Research for the grant MASF\_FFABR\_17\_01. This investigation is supported by the National Aeronautics and Space Administration (NASA) grants GO9-20083X and GO0-21110X. In this work we extensively used \texttt{TOPCAT}\footnote{\url{http://www.star.bris.ac.uk/~mbt/topcat/}} \citep{taylor2005topcat} and \texttt{astropy}\footnote{\url{https://www.astropy.org/index.html}} \citep{astropy2013A&A...558A..33A,astropy2018AJ....156..123A} for preparation and manipulation of the data.

The \textit{Fermi} LAT Collaboration acknowledges generous ongoing support
from a number of agencies and institutes that have supported both the
development and the operation of the LAT as well as scientific data analysis.
These include the National Aeronautics and Space Administration and the
Department of Energy in the United States, the Commissariat \`a l'Energie Atomique
and the Centre National de la Recherche Scientifique / Institut National de Physique
Nucl\'eaire et de Physique des Particules in France, the Agenzia Spaziale Italiana
and the Istituto Nazionale di Fisica Nucleare in Italy, the Ministry of Education,
Culture, Sports, Science and Technology (MEXT), High Energy Accelerator Research
Organization (KEK) and Japan Aerospace Exploration Agency (JAXA) in Japan, and
the K.~A.~Wallenberg Foundation, the Swedish Research Council and the
Swedish National Space Board in Sweden.
 
Additional support for science analysis during the operations phase is gratefully
acknowledged from the Istituto Nazionale di Astrofisica in Italy and the Centre
National d'\'Etudes Spatiales in France. This work performed in part under DOE
Contract DE-AC02-76SF00515.

%\appendix

%\section{A}

%\label{ap:tables}

%TBD

%%%%%%%%%%%%%%%%%%%%%%%%%%%%%%%%%%%%%%%%%%%%%%%%%%

\bibliography{refs}{}
\bibliographystyle{aasjournal}



\end{document}